\documentclass[reprint,twocolumn,amsmath,showpacs,superscriptaddress,amssymb,aps,prb]{revtex4-1}

\usepackage{graphicx}% Include figure files
\usepackage{graphics}
\usepackage{amsmath}
\usepackage{dcolumn}% Align table columns on decimal point
\usepackage{bm}% bold math

\begin{document}

%\preprint{APS/123-QED}

\title{Two-Dimensional Wide-Band-Gap II-V Semiconductors with a Dilated Graphene-like Structure}

\author{Xue-Jing Zhang}
\affiliation{Beijing National Laboratory for Condensed Matter Physics, Institute of Physics, Chinese Academy of Sciences, Beijing 100190, China}%
\affiliation{School of Physical Sciences, University of Chinese Academy of Sciences, Beijing 100190, China}
\author{Bang-Gui Liu}\email{bgliu@iphy.ac.cn}
\affiliation{Beijing National Laboratory for Condensed Matter Physics, Institute of Physics, Chinese Academy of Sciences, Beijing 100190, China}%
\affiliation{School of Physical Sciences, University of Chinese Academy of Sciences, Beijing 100190, China}

\date{\today}% It is always \today, today, but any date may be explicitly specified

\begin{abstract}
Since the advent of graphene, two-dimensional (2D) materials become very attractive and there is growing interest to explore new 2D beyond graphene. Here,  through density functional theory (DFT) calculations, we predict 2D wide-band-gap II-V semiconductor materials of M$_3$X$_2$ (M=Zn, Cd and X=N, P, As) with a dilated graphene-like honeycomb structure. The structure features that the group-V X atoms form two X-atomic planes symmetrically astride the centering group-IIB M atomic plane. The 2D Zn$_3$N$_2$, Zn$_3$P$_2$, and Zn$_3$As$_2$ are shown to have direct band gaps of 2.87, 3.81, and 3.55 eV, respectively, and the 2D Cd$_3$N$_2$, Cd$_3$P$_2$, and Cd$_3$As$_2$ exhibit indirect band gaps of 2.74, 3.51, and 3.29 eV, respectively. Each of the six 2D materials is shown to have effective carrier (either hole or electron) masses down to $0.03\sim 0.05$ $m_0$. The structural stability and feasibility of experimental realization of these 2D materials has been shown in terms of DFT phonon spectra and total energy comparison with related existing bulk materials. On the experimental side, there already are many similar two-coordinate structures of Zn and other transition metals in various organic materials, which can be considered to support our DFT prediction. Therefore, these 2D semiconductors can enrich the family of 2D electronic materials and may have promising potential for achieving novel transistors and optoelectronic devices.
\end{abstract}

% PACS, the Physics and Astronomy Classification Scheme.

%\keywords{Suggested keywords}%Use showkeys class option if keyword display desired

\maketitle

%\tableofcontents

\section{Introduction}

Since graphene was created, two-dimensional (2D) materials have been attracting a lot of attention due to their unique and exceptionally important properties and promising applications in next-generation electronic, optoelectronic, spintronic, and other novel devices\cite{1,2,3,4,5,6,6a,7,7a}. Graphene, with massless Dirac Fermions weakly sensitive to backscattering and travelling at very high speed over very wide distances at room temperature\cite{6}, has superior performance, but its zero-gap behavior is an obstacle to switch current on and off in field-effect transistors (FETs), being inconsistent with modern semiconductor technology\cite{3,4,7,7a}. As for other 2D compounds, h-BN has a wide gap of 5.5 eV\cite{8} and on the other hand, MoS$_2$ monolayer, black phosphorene, and As monolayer have band gaps of 1.8, 1.51, and 2.49 eV\cite{9,10,11}, respectively. Monolayer MoS$_2$ FET have been demonstrated with carrier mobility of 200 cm$^2$V$^{-1}$s$^{-1}$ at room temperature, current on/off ratios of $1\times 10^8$, and ultralow standby power dissipation\cite{12}. Black P, arranged in a honeycomb puckered lattice, presents a high electron mobility of 1000 cm$^2$V$^{-1}$s$^{-1}$ and high in-plane anisotropy\cite{13,14}. These 2D materials can be applied in nanoelectronic devices. However, the 2D semiconductors with band gaps of $2.5\sim 5.0$ eV are highly desirable for rich applications in electronic and optoelectronic devices.

Here, we predict six 2D II-V wide-band-gap semiconductors M$_3$X$_2$ (M=Zn, Cd and X=N, P, As) through systematical density functional theory (DFT) calculations. The dilated graphene-like honeycomb structures of the 2D M$_3$X$_2$ (M=Zn, Cd and X=N, P, As) are formed according to normal valence rules that each group-IIB M atom is surrounded by two group-V X atoms and each X is surrounded by three M atoms. The structural stability and feasibility of experimental realization of these 2D materials has been investigated in terms of DFT structural optimization, phonon spectra, and total energy comparison with related existing 3D materials. On the experimental side, similar two-coordinate structures of metallic atoms have been observed in various organic materials\cite{32,33}. Our DFT study shows that these 2D semiconductors have wide band gaps ranging from 2.74 to 3.81 eV and most of their effective carrier masses are from 0.03 to 0.05 $m_0$. They may be potential materials for achieving novel transistors with high on/off ratios and optoelectronic devices working under violet or UV light. More detailed results will be presented in the following.

\begin{figure*}[!tbp]
\centering  % Requires \usepackage{graphicx}
\includegraphics[width=12cm]{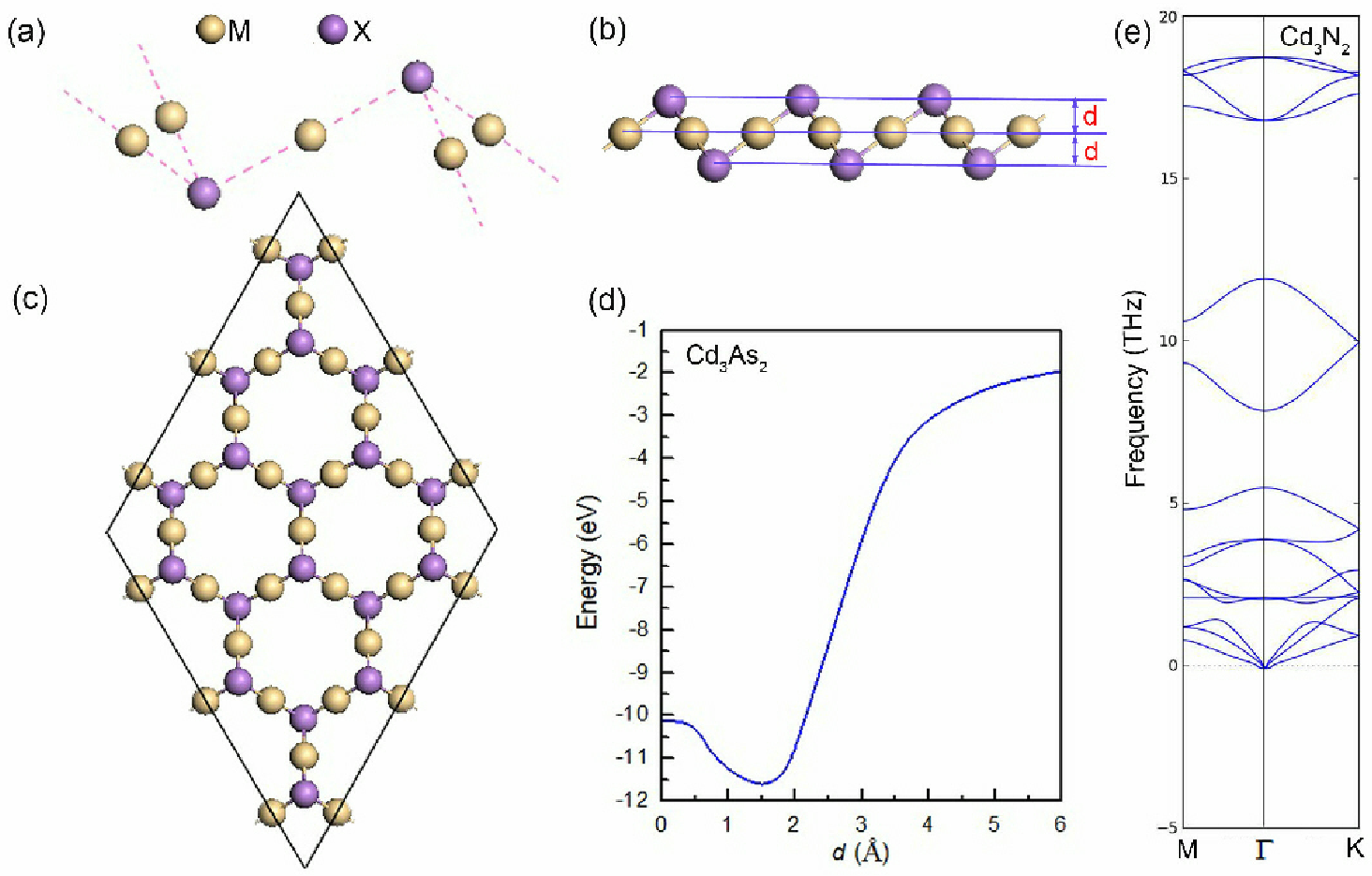}\\
\caption{(Color online) (a) Crystal structure of the 2D II-V materials of M$_3$X$_2$ (M=Zn, Cd and X=N, P, As). (b, c) Side and top views of the buckled 2D M$_3$X$_2$ (M=Zn, Cd and X=N, P, As). (d) The energy of the 2D Cd$_3$As$_2$ with the largest buckling as a function of buckling parameter $d$. (e) Phonon spectra of the 2D Cd$_3$N$_2$ as the representative.}\label{fig1}
\end{figure*}

\section{Computational Details}

Our first-principles calculations are performed with the projector-augmented wave method as implemented in the Vienna {\it Ab-initio} Simulation Package (VASP)\cite{15,16}. The valence-electron configuration of group-IIB M and group-V X are d$^{10}$s$^2$ and s$^2$p$^3$, respectively. We use a slab model to simulate the 2D materials by adding a 15 \AA{} thick vacuum layer to ensure the decoupling between different tri-layers (consisting of one M layer and two X layers) in the calculational model. A $\Gamma$-centered $9\times 9\times 1$ $k$-grid is employed and the plane wave energy cutoff is set to 500 eV. Furthermore, the convergence criterium for the total energy is chosen to be 10$^{-5}$ eV. The cell geometry and the ionic positions are both optimized until the Hellmann-Feynmann forces on each atom become less than 0.01 eV/\AA{}. For structure optimization, we use the generalized gradient approximation (GGA) by Perdew, Burke and Ernzerhof (PBE)\cite{17} for the exchange-correlation potential. The electronic structures (DOSs and bands) are calculated by both PBE and hybrid functional HSE06.\cite{18} The latter is used to improve the theoretical description of the electronic structures, especially the semiconductor gaps. Phonon spectra are calculated based on Phonopy\cite{19,20}.

\section{Results and discussion}

\subsection{Structures and stability}

For constructing reasonable 2D structures for M$_3$X$_2$ (M=Zn, Cd and X=N, P, As), the valences of group-IIB M and group-V X must be taken into account. First, we can assume that M and X are in the same plane and form a 2D dilated-honeycomb structure (space group P6/mmm), in which each M has two X neighbors and each X has three M ones. This structure has the same rotational symmetry as grapheme, but the C atom at the hexagonal vertex is replaced by X atom and there is an M atom at the middle of each dilated hexagonal edge of the honeycomb unit. There are six M and six X atoms in the hexagon of the honeycomb unit. Unfortunately, this structure is unstable against a buckling of X atoms from the plane. The buckling, keeping the three-fold rotational symmetry, makes two X planes remain symmetrically astride the M plane and it can be parameterized with the distance $d$ between the M plane and both of the two X planes. This buckling makes the space group reduce to P$\bar{3}$m1, but lowers the total energy by 0.042, 1.027, 1.419, 0.215, 1.076, and 1.476 eV per formula unit for the Zn$_3$N$_2$, Zn$_3$P$_2$, Zn$_3$As$_2$, Cd$_3$N$_2$, Cd$_3$P$_2$, and Cd$_3$As$_2$, respectively. With the optimized $d$ value and the 2D crystal lattice constant $a$, we can calculate the M-X bond length $l_{\rm MX}$. We summarize $a$, $d$, and $l_{\rm MX}$ values in Table I, and present the 2D crystal structure in Figs. 1(a)-(c). It is clear that this 2D structure can be seen as a dilated, buckled honeycomb structure like graphene. To better demonstrate the effect of the buckling, we present the total energy of the 2D Cd$_3$As$_2$ as a function of $d$ in Fig. 1(d) because of the largest buckling in this case. Furthermore, we have calculated the phonon spectra in all the six cases. The phonon spectra of the 2D Cd$_3$N$_2$ as the representative are presented in Fig. 1(e). Because there is no soft phonon mode, there is no kinetic instability for any of the six 2D M$_3$X$_2$ (M=Zn, Cd and X=N, P, As) materials\cite{21}.

\begin{table}[!h]
\caption{Calculated 2D lattice constant $a$, buckling parameter $d$, and M-X bond length $l_{\rm MX}$ of the 2D II-V materials of M$_3$X$_2$ (M=Zn, Cd and X=N, P, As). Also presented are calculated energy difference values per atom, $\Delta E$, between the 2D M$_3$X$_2$ materials and the 3D bulk M$_3$X$_2$ with space group  Ia$\bar{3}$ (for bulk Zn$_3$N$_2$ and Cd$_3$N$_2$)\cite{22,23,24} and space groups P4$_2$/nmc and P4$_2$32 (for bulk M$_3$X$_2$ with M=Zn, Cd and X=P, As)\cite{25,26,27,28,29,30,31}.}
\begin{ruledtabular}
\begin{tabular}{ccccc}
System &	$a$ (\AA) &	$d$ (\AA)& $l_{\rm MX}$ (\AA) &	$\Delta E$ (eV)\\ \hline
Zn$_3$N$_2$ &	6.157 &	0.427 &	1.828 &	0.207 \\
Zn$_3$P$_2$ &	6.418 &	1.291 &	2.258 &	0.259, 0.075 \\
Zn$_3$As$_2$ & 6.484 &1.453 &	2.370 &	0.211, 0.054 \\
Cd$_3$N$_2$ &	6.745 &	0.628 &	2.063 &	0.207 \\
Cd$_3$P$_2$ &	7.121 &	1.344 &	2.456 &	0.219, 0.029 \\
Cd$_3$As$_2$ & 7.185 &1.505 &	2.563 &	0.185, 0.032
\end{tabular}
\end{ruledtabular}
\end{table}

For analyzing the energy stability of the 2D M$_3$X$_2$ (M=Zn, Cd and X=N, P, As) and their feasibility of being experimentally synthesized, we need to compare their total energies with those of corresponding 3D bulk materials that have already been experimentally realized, in addition to checking the phonon spectra. These 3D bulk materials assume different crystal structures with space groups: Ia$\bar{3}$ (for Zn$_3$N$_2$ and Cd$_3$N$_2$)\cite{22,23,24}, P4$_2$/nmc (for Zn$_3$P$_2$, Cd$_3$P$_2$, Zn$_3$As$_2$ and Cd$_3$As$_2$) \cite{25,26,27,28,29,30}, and P4$_2$32 (for Zn$_3$P$_2$, Cd$_3$P$_2$, Zn$_3$As$_2$ and Cd$_3$As$_2$)\cite{31}. We have calculated the energy differences per atom, $\Delta E$, between the 2D M$_3$X$_2$ (M=Zn, Cd and X=N, P, As) and the corresponding 3D crystal materials, and present the calculated results in Table I. As for both Zn$_3$N$_2$ and Cd$_3$N$_2$, the $\Delta E$ value is an energy difference per atom of 0.21 eV. For both Cd$_3$P$_2$ and Cd$_3$As$_2$, the energy difference per atom is 0.03 eV, the smallest. The largest value for the energy difference per atom is 0.26 in the case of Zn$_3$P$_2$. Therefore, these small energy differences, ranging from 0.03 eV to 0.26 eV, with the phonon spectra free of any soft mode, imply that these 2D materials could be realized experimentally.

\begin{figure}[!tbp]
\centering  % Requires \usepackage{graphicx}
\includegraphics[clip, width=8.4cm]{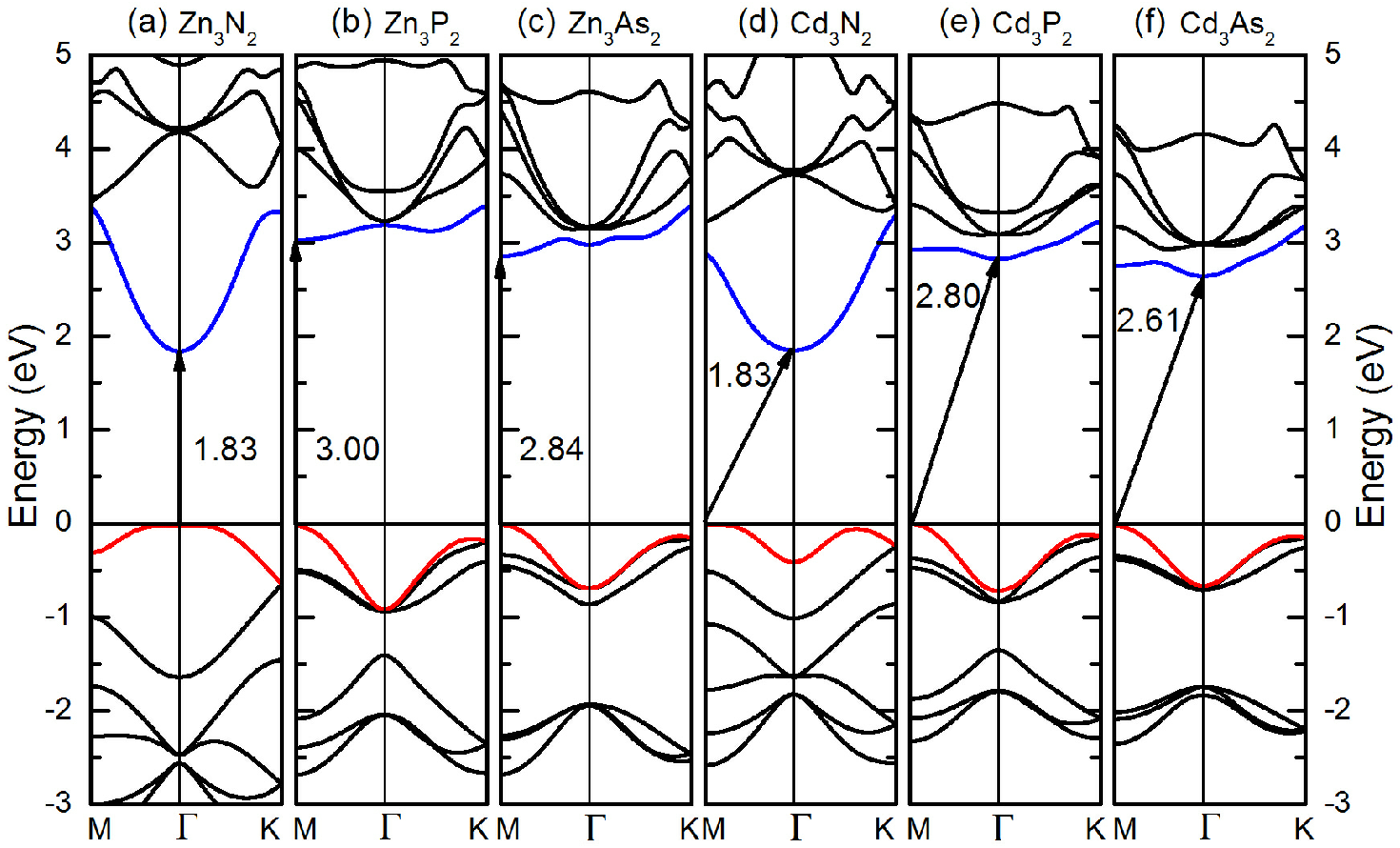}\\
\caption{(Color online) The band structures ($E_F = 0$ eV) calculated with PBE\cite{17} for the six 2D semiconductors of M$_3$X$_2$ (M=Zn, Cd and X=N, P, As).}
\end{figure}

\begin{figure}[!tbp]
\centering  % Requires \usepackage{graphicx}
\includegraphics[clip, width=8.4cm]{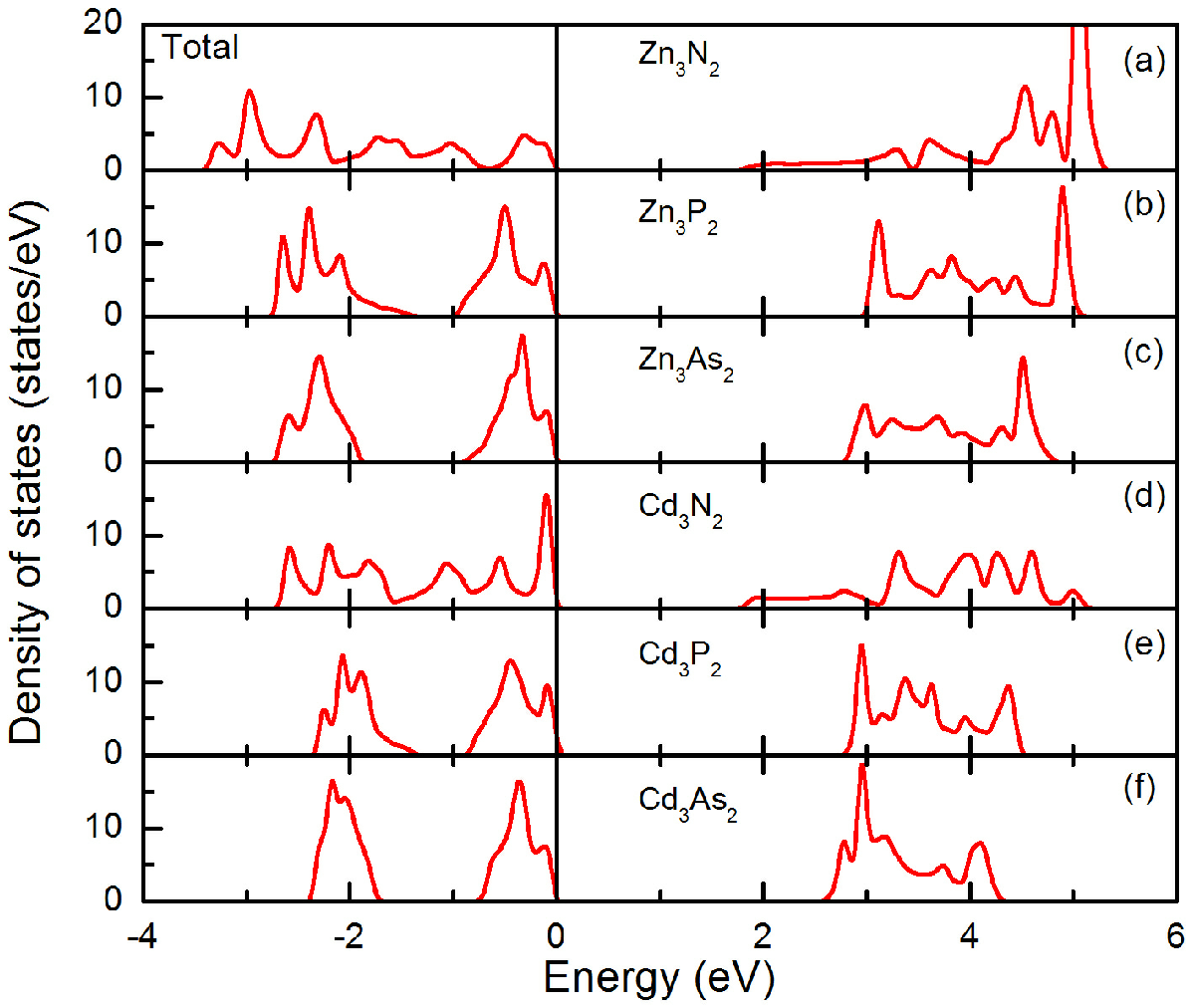}\\
\caption{(Color online) The total DOS ($E_F = 0$ eV) calculated with PBE\cite{17} of the six 2D II-V materials of M$_3$X$_2$ (M=Zn, Cd and X=N, P, As).}
\end{figure}

\subsection{Electronic structures}

We present in Fig. 2 the electronic band structures of the 2D M$_3$X$_2$ (M=Zn, Cd and X=N, P, As) calculated with PBE. It is clear that the Zn$_3$N$_2$ is a direct band gap semiconductor with a band gap of 1.83 eV at $\Gamma$ point of the Brillouin zone, while the Zn$_3$P$_2$ and the Zn$_3$As$_2$ are direct band gap semiconductors with band gaps of 3.00 and 2.84 eV at M point. However, the 2D Cd$_3$X$_2$ (X=N, P, As) are indirect band gap semiconductors with band gaps of 1.83, 2.80 and 2.61 eV, and their valence band maxima (VBM) are at M point and their conduction band minima (CBM) at $\Gamma$ point. The total DOSs of the 2D M$_3$X$_2$ (M=Zn, Cd and X=N, P, As) are presented in Fig. 3. The energy distribution of total DOS varies with different anion X (X=N, P, As). The band gap of the Zn$_3$X$_2$ (X=N, P, or As) is slightly larger than that of the corresponding Cd$_3$X$_2$, respectively.

Fig. 4 shows the partial DOSs of N and Zn in the 2D Zn$_3$N$_2$, those of N and Cd in the 2D Cd$_3$N$_2$, and those of As and Cd in the 2D Cd$_3$As$_2$. It can be seen that the VBM of these M$_3$X$_2$ materials arise from the X p orbitals. For the 2D Zn$_3$N$_2$, N p$_z$ state dominates between -1.7 eV and 0 eV and the p$_x$ and p$_y$ states hybrid with Zn d and s states in the window from -3.42 eV to -1.38 eV. The N-p dominant bands in the Cd$_3$N$_2$ are more narrow than those in the Zn$_3$N$_2$. Near the VBM (from -0.25 eV to 0 eV) of the 2D Cd$_3$N$_2$, there is a strong peak of the p$_z$ state with small p$_x$ and p$_y$ states admixtures. The N p$_x$ and p$_y$ states hybrid with Cd s and d, as demonstrated in the energy window from -2.7 eV to -1.6 eV. In the 2D Cd$_3$As$_2$, As p states dominate between -0.8 and 0 eV and hybrid with Cd s states (Cd d states are too small to need to be shown here) between -2.3 and -1.7 eV. Our calculations also show that the partial DOSs of the 2D Zn$_3$P$_2$, Cd$_3$P$_2$, and Zn$_3$As$_2$ are similar to those of the 2D Cd$_3$As$_2$.

\begin{figure*}[!tbp]
\centering  % Requires \usepackage{graphicx}
\includegraphics[clip, width=12cm]{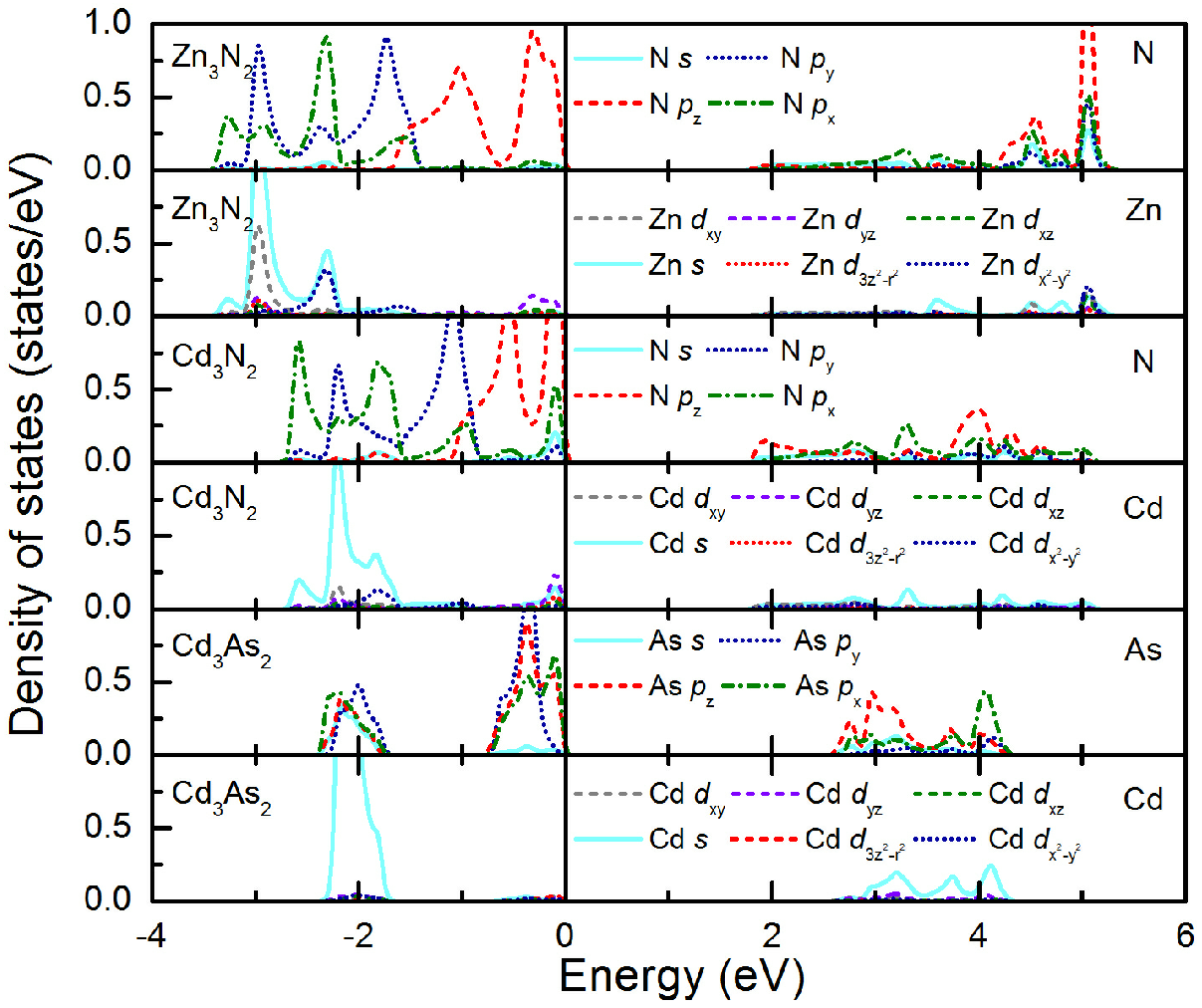}\\
\caption{(Color online) The partial DOS ($E_F = 0$ eV, calculated with GGA, and projected in the atomic muffin tins) of the three 2D II-V semiconductor materials: Zn$_3$N$_2$, Cd$_3$N$_2$, and Cd$_3$As$_2$. It should be pointed out that the partial DOS of the interstitial region is not included here. }
\end{figure*}

\begin{figure*}[!tbp]
\centering  % Requires \usepackage{graphicx}
\includegraphics[clip, width=12cm]{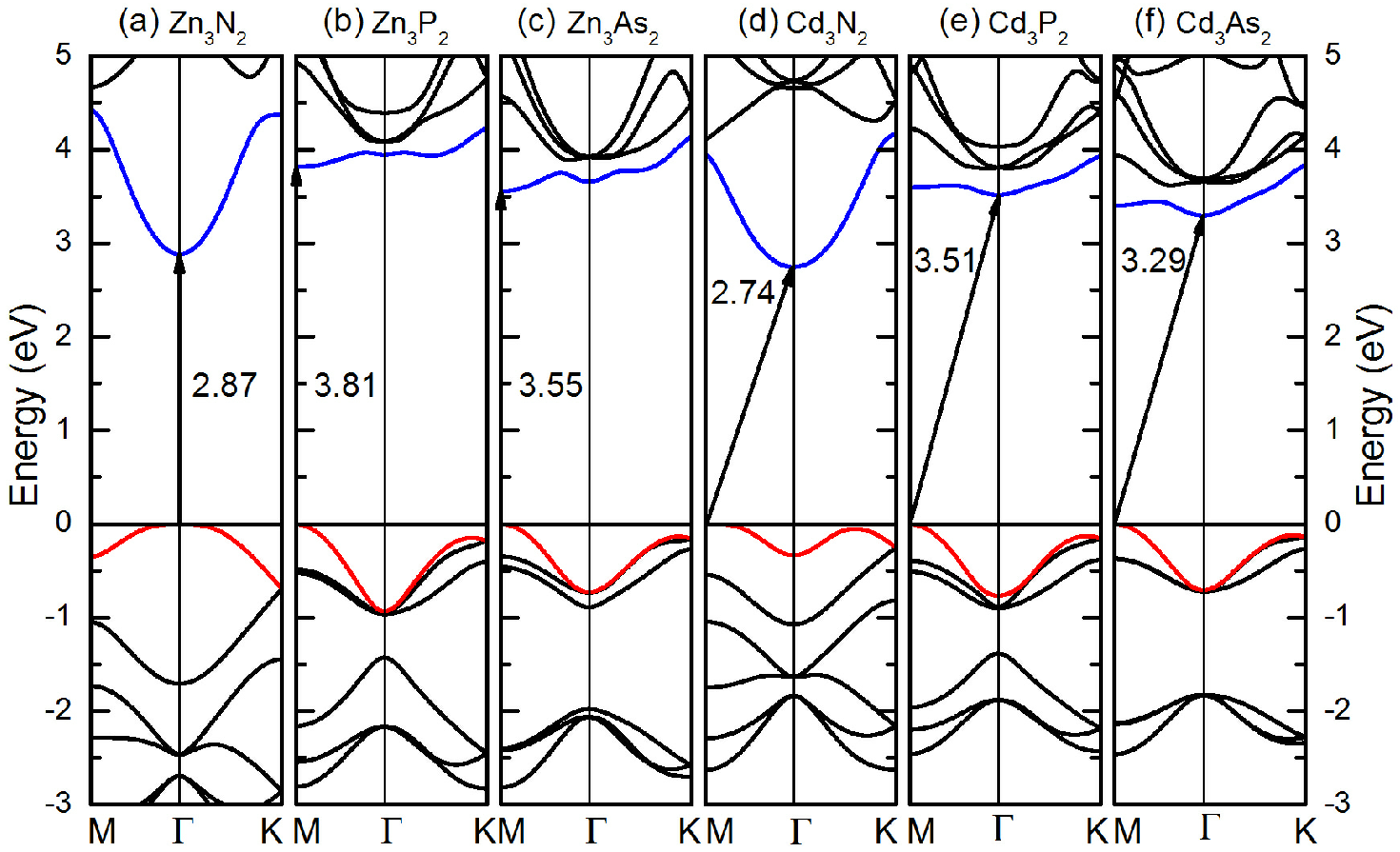}\\
\caption{(Color online) The band structures ($E_F= 0$ eV) calculated with HSE06 functional\cite{18} of the six 2D materials of M$_3$X$_2$ (M=Zn, Cd and X=N, P, As), with the gaps being indicated.}
\end{figure*}

\subsection{Improved electronic properties}

It is well known that GGA usually underestimates semiconductor gaps. We use HSE06 functional\cite{18} to achieve a better description of band gaps in the 2D semiconductors. The electronic band structures of the six 2D M$_3$X$_2$ (M=Zn, Cd and X=N, P, As) calculated with HSE06 are presented in Fig. 5. Clearly, HSE06 functional leads to the wider band gaps of 2.87, 3.81, 3.55, 2.74, 3.51 and 3.29 eV than those with PBE for the 2D Zn$_3$N$_2$, Zn$_3$P$_2$, Zn$_3$As$_2$, Cd$_3$N$_2$, Cd$_3$P$_2$, and Cd$_3$As$_2$, respectively. These band gaps are wider than those of successfully fabricated MoS$_2$ and recently theoretically studied As monolayer\cite{11,34}.

Using HSE06 hybrid functional, we have also calculated effective carrier masses $m^*$ of the 2D M$_3$X$_2$ (M=Zn, Cd and X=N, P, As) at VBM and CBM. The calculated results are summarized in Table II. For the 2D Zn$_3$N$_2$, the calculated hole effective masses $m^*_{\rm \Gamma K}$  and $m^*_{\rm \Gamma M}$  are 0.24$m_0$ and 0.26$m_0$ ($m_0$ is the mass of free-electron), which are close to those of arsenene ($m^*_{\rm \Gamma K}$=0.23$m_0$, $m^*_{\rm \Gamma M}$=0.29$m_0$) and antimonene ( $m^*_{\rm \Gamma K}$=0.20$m_0$,  $m^*_{\rm \Gamma M}$=0.24$m_0$)\cite{11} and smaller than those of monolayer MoS$_2$ ($m^*$=0.48$m_0$)\cite{35}. For the 2D Cd$_3$N$_2$, the calculated hole effective mass $m^*_{\rm M\Gamma}$  is 0.37$m_0$. However, For the 2D Zn$_3$P$_2$, Zn$_3$As$_2$, Cd$_3$P$_2$ and Cd$_3$As$_2$, the hole effective masses $m^*_{\rm M\Gamma}$  are 0.04$m_0$, 0.04$m_0$, 0.03$m_0$ and 0.03$m_0$, maybe being able to cause high carrier mobilities. For the Zn$_3$N$_2$ and Cd$_3$N$_2$, the calculated electron effective masses, ~0.03$m_0$, are much smaller than the hole effective masses, but for the 2D Zn$_3$P$_2$ and Zn$_3$As$_2$, the electron effective masses, 0.17 and 0.10$m_0$, are substantially larger than the hole effective mass values. As for 2D Cd$_3$P$_2$ and Cd$_3$As$_2$, the calculated electron effective masses, 0.05 and 0.04$m_0$, are only slightly larger than the hole effective masses.

\begin{table}[!htbp]
\caption{Semiconductor gaps (eV) and effective carrier mass values calculated with HSE06\cite{18} of the six 2D M$_3$X$_2$ (M=Zn, Cd and X=N, P, As). The band gap values in parentheses are calculated with PBE.}
\begin{ruledtabular}
\begin{tabular}{cccc}
System	& 	     Gap & 	  Hole mass &	Electron mass \\ \hline
Zn$_3$N$_2$ &  2.87(1.83) &	0.24$m_0$($\Gamma$K),0.26$m_0$($\Gamma$M) &	0.03$m_0$\\
Zn$_3$P$_2$	&  3.81(3.00) &	0.04$m_0$(M$\Gamma$) &	0.17$m_0$(M$\Gamma$) \\
Zn$_3$As$_2$ & 3.55(2.84) & 0.04$m_0$(M$\Gamma$) & 0.10$m_0$(M$\Gamma$) \\
Cd$_3$N$_2$ &  2.74(1.83) & 0.37$m_0$(M$\Gamma$) & 0.03$m_0$\\
Cd$_3$P$_2$ &  3.51(2.80) & 0.03$m_0$(M$\Gamma$) & 0.05$m_0$ \\
Cd$_3$As$_2$ & 3.29(2.61) & 0.03$m_0$(M$\Gamma$) & 0.04$m_0$
\end{tabular}
\end{ruledtabular}
\end{table}

\subsection{Further Discussion}

The cation (Zn or Cd) has a nominal valence of 2+ and the anion (N, P, or As) a nominal valence of 3-. Actually, the bond is not purely ionic and part of it is covalent, which can be shown by the nonzero weight of the filled Zn (or Cd) s state in Fig. 4. Consequently, the valence bands in Fig. 3 are originated mainly from the p states of the anion and the conduction bands mainly from the s states of the cation. This can be seen through comparing Fig. 3 and Fig. 4 that the partial DOSs (projected in the atomic muffin tins) of the conduction bands are substantially smaller than those of the valence bands, and therefore the wave functions of the conduction bands are mainly in the interstitial region, being more extended in the real space. The incomplete transferring of the cation s electrons in the 2D II-V semiconductors is partly because the cation has only two coordinates and partly because the bond includes some covalence as it does in usual II-VI semiconductors.

At this stage, however, one may ask such a question: Can the 2D II-V semiconductors including two-coordinate cations (Zinc or Cadmium) be experimentally realized? On the theoretical side, we have shown through DFT calculated phonon spectra and total energy comparison with existing 3D materials that these 2D II-V semiconductor materials are structurally stable and should be experimentally realizable, although the cation has only two coordinates. On the experimental side, it has been already shown that one Zinc (and other transition metal) atom can indeed have two coordinates in various organic materials including Zinc (and other transition metals)\cite{32,33}. Therefore, the answer is certain.

The semiconductor gaps in these 2D semiconductors are between that of h-BN and those of MoS$_2$, black P, and As monolayer\cite{8,9,10,11}. The effective hole masses of the 2D Zn$_3$N$_2$ and Cd$_3$N$_2$ are slightly less than those of 2D MoS$_2$, antimonene, and arsenene, but most of the effective carrier (either hole or electron) masses range from 0.03 to 0.05 $m_0$.\cite{11,33} Their 2D dilated honeycomb structure makes them be structurally compatible with most of the promising 2D materials, and thus they can be used to form hybrid structures with other 2D materials in order to design novel electronic, spintronic, optoelectronic, and other devices\cite{3,4,5,6,6a,7,7a}.

\section{Conclusion}

In summary, we have predicted the six two-dimensional (2D) wide-band-gap II-V semiconductor materials M$_3$X$_2$ (M=Zn, Cd and X=N, P, As) with a dilated, buckled graphene-like honeycomb structure. The structural stability and feasibility of experimental realization of these 2D materials has been shown through DFT phonon spectra and total energy comparison with related existing 3D materials. They are also supported by the fact that there have already been many similar two-coordinate structures of Zn and other metallic atoms in various organic materials\cite{32,33}. Our DFT study has shown that the Zn$_3$N$_2$, Zn$_3$P$_2$, and Zn$_3$As$_2$ are direct band gap semiconductors with band gaps of 2.87, 3.81, and 3.55 eV, and however, the Cd$_3$N$_2$, Cd$_3$P$_2$, and Cd$_3$As$_2$ are indirect band gap semiconductors with band gaps of 2.74, 3.51, and 3.29 eV, respectively. The dilated, buckled honeycomb structure enriches the family of two-dimensional materials. Our further DFT calculations show that the effective hole masses of the 2D Zn$_3$N$_2$ and Cd$_3$N$_2$ are slightly less than that of well-known 2D MoS$_2$, and each of the six 2D materials has effective carrier (either hole or electron) masses down to 0.03 to 0.05 $m_0$, having potential to achieve high carrier mobility. Therefore, these 2D II-V semiconductors may have promising applications in future electronic and optoelectronic devices.

\begin{acknowledgments}
This work is supported by the Nature Science Foundation of China (Grant No. 11174359 and No. 11574366), by the Department of Science and Technology of China (Grant No. 2016YFA0300701 and No. 2012CB932302), and by the Strategic Priority Research Program of the Chinese Academy of Sciences (Grant No.XDB07000000).
\end{acknowledgments}

\end{document}